\documentstyle[aps,prl,epsf,preprint]{revtex}
\begin{document}
\draft
\title{Magnetic Excitations in the Spin Gap System TlCuCl$_3$} 

\author{A. Oosawa,$^{1}$ T. Kato,$^{2}$ H. Tanaka,$^{1}$ K. Kakurai,$^{3,*}$
\ \\
M. M\"{u}ller$^{4}$ and H. -J. Mikeska$^{4}$}

\address{
$^{1}$Department of Physics, Tokyo Institute of Technology, Meguro-ku, Tokyo 152-8551, Japan \\
$^{2}$Faculty of Education, Chiba University, Inage-ku, Chiba 263-8522, Japan \\
$^{3}$Advanced Science Research Center, Japan Atomic Energy Research Institute,
Tokai, Ibaraki 319-1195, JAPAN \\
$^{4}$Institut f\"{u}r Theoretische Physik, Universit\"{a}t Hannover, Appelstrasse 2, 30167 Hannover, Germany \\
}

\maketitle

\begin{abstract}
TlCuCl$_3$ has a singlet ground state with the excitation gap
$\Delta =0.65$ meV. The magnetic excitations in TlCuCl$_3$ have been
investigated by means of neutron inelastic scattering experiments. The
constant-$\it{\pmb Q}$ energy scan profiles were collected in the $a^*-c^*$ plane. A well-defined single magnetic excitation mode was observed. The dispersion relations along four different directions were
determined. The lowest excitation occurs at ${\it{\pmb Q}}=(h, 0, l)$ with
integer $h$ and odd $l$, as observed in KCuCl$_3$. A cluster series
expansion up to the sixth order was applied to analyze the dispersion
relations, and the individual exchange interactions were evaluated. It
was demonstrated that TlCuCl$_3$ is a strongly coupled spin-dimer
system.
\end{abstract}
$^*$Present Address: Advanced Science Research Center, Japan Atomic Energy Research Institute, Tokai, Ibaraki 319-1195, JAPAN

\section{INTRODUCTION}
The magnetic excitations in quantum spin systems, which have a singlet
ground state with an excitation gap, are a new
subject in magnetism. Recently, neutron inelastic scattering
experiments have been actively performed to investigate the magnetic
excitations in various coupled spin-dimer systems, {\it e.g.},
(VO)$_2$P$_2$O$_7$ \cite{Garrett}, Cu(NO$_3$)$_2 \cdot
\frac{5}{2}$D$_2$O \cite{Xu}, BaCuSi$_2$O$_6$ \cite{Sasago},
SrCu$_2$(BO$_3$)$_2$ \cite{Kageyama} and KCuCl$_3$
\cite{Kato1,Cavadini1,Kato2,Cavadini2}. A variety of dispersion
relations for the propagation of the excited triplet have been
observed in these systems, and the true nature of the exchange
networks, which was not expected from the crystal structures, has been
revealed in (VO)$_2$P$_2$O$_7$ and KCuCl$_3$.

Since large single crystals of KCuCl$_3$ can be obtained, its magnetic excitations have been extensively investigated by neutron inelastic scattering \cite{Kato1,Cavadini1,Kato2,Cavadini2,Cavadini3,Kato3}. The dispersion relations were first analyzed using the effective dimer approximation, in which the exchange interactions between individual dimers are reduced to an effective interaction between dimers \cite{Cavadini1,Cavadini2,Kato3,Suzuki}. Later, the cluster series expansion was applied by M\"{u}ller and Mikeska \cite{Mueller} to describe the dispersion relation in order to evaluate the individual exchange interactions. From these analyses, the exchange network in KCuCl$_3$ was elucidated. Consequently, KCuCl$_3$ has been characterized as a weakly and three-dimensionally coupled spin-dimer system.

In this study, we investigate the magnetic excitations in TlCuCl$_3$, which is isostructural with KCuCl$_3$ \cite{Takatsu}. This compound has a monoclinic structure (space group $P2_1/c$) \cite{Willett}. The crystal structure is composed of planar dimers of Cu$_2$Cl$_6$, which are stacked on top of one another to form infinite double chains parallel to the crystallographic $a$-axis. These double chains are located at the corners and center of the unit cell in the $b-c$ plane, and are separated by Tl$^+$ ions. Figure 1 shows the projection of Cu$^{2+}$ ions with spin-$\frac{1}{2}$ on the $a-c$ plane and the exchange interactions between Cu$^{2+}$ ions. For the notation of the exchange interaction, see reference \cite{Mueller}.

The magnetic ground state of TlCuCl$_3$ is a spin singlet with an excitation gap, as observed in KCuCl$_3$ \cite{Takatsu}. The magnitude of the spin gap $\Delta$ in TlCuCl$_3$ was evaluated from the critical field of the magnetization curve \cite{Shiramura1,Oosawa1} and the excitation energy of the direct ESR transition \cite{Tanaka1} as $\Delta =0.65$ meV. The lattice parameters, the critical fields and the saturation fields for KCuCl$_3$ and TlCuCl$_3$ are listed in Table I. The crystal lattice of TlCuCl$_3$ is compressed along the $a$-axis and enlarged in the $b-c$ plane as compared with KCuCl$_3$. Thus, substituting Tl$^{+}$ for K$^{+}$ produces uniaxial stress along the $a$-axis.

Although the crystal structures of TlCuCl$_3$ and KCuCl$_3$ are the
same, there is a significant difference between their magnetic
properties, $i.e.$, the spin gap for TlCuCl$_3$ is one-quarter of that
for KCuCl$_3$, while the saturation field for TlCuCl$_3$ is about
twice as large as that for KCuCl$_3$ \cite{Shiramura1,Tatani}. This
suggests that the interdimer interactions in TlCuCl$_3$ are much
stronger than those in KCuCl$_3$. However, the details of the individual exchange interactions in TlCuCl$_3$ have not been clarified so far. 

Recently the field-induced magnetic phase transition was observed by magnetization and specific heat measurements \cite{Oosawa1,Oosawa3}. It was demonstrated that the nature of the phase transition can be described in terms of the Bose-Einstein condensation (BEC) of the triplet excitations (magnons) \cite{Nikuni}. In the magnon BEC theory, the magnons around the lowest excitation are relevant to the phase transition. For these reasons, we have investigated the magnetic excitations in TlCuCl$_{3}$.

Recently Cavadini {\it et al.} \cite{Cavadini4} have also independently investigated the magnetic excitations in TlCuCl$_3$ by means of neutron inelastic scattering. They measured mainly the dispersions parallel to the principal axes $a^*$, $b^*$ and $c^*$, and evaluated the intradimer exchange interaction and the effective interactions between dimers, which are given by the linear combinations of the individual exchange interactions as shown in section III. Although the present measurements are confined in the $a^*c^*$-plane, we measured the dispersion relations not only parallel to the $a^*$ and $c^*$-axes, but also for two diagonal directions $(h, 0, 2h+1)$ and $(h, 0, -2h+1.4)$, which are roughly perpendicular to each other. As shown in the next section, the magnetic excitation is most dispersive along $(h, 0, 2h+1)$, while it is less dispersive along $(h, 0, -2h+1.4)$. The $(h, 0, 2h+1)$ direction is parallel to the cleavage $(1, 0, -2)$ plane, in which the hole orbitals of Cu$^{2+}$ spread. The dispersion relations along $(h, 0, 2h+1)$ and $(h, 0, -2h+1.4)$ are essential, because the interdimer exchange interactions cannot be uniquely determined without them, {\it i.e.}, there is another set of exchange parameters which can fit the dispersion relations principal axes $a^*$, $b^*$ and $c^*$. We analyzed the dispersion relations using a cluster series expansion to sixth order to evaluate the individual interdimer exchange interactions, which cannot be determined within the framework of the effective dimer approximation \cite{Cavadini4}. The effective interactions calculated with these individual interdimer exchange interactions are significantly different from those obtained by effective dimer approximation.

The arrangement of this paper is as follows. In the next section, experimental details are described, and the experimental results are presented. In section III, the experimental results are supplemented by a theoretical analysis based on dispersion curves calculated from a cluster series expansion, and the individual exchange interactions are determined. Section IV is devoted to conclusions.

\section{EXPERIMENTS AND RESULTS}       
TlCuCl$_3$ single crystals were grown from a melt by the Bridgman method. The details of sample preparation have been reported in reference \cite{Oosawa1}. The TlCuCl$_3$ crystal cleaves easily parallel to the (0, 1, 0) plane, in which lie the $a^*$- and $c^*$-axes.

Neutron inelastic scattering was performed using the ISSP-PONTA spectrometer installed at JRR-3M, Tokai. The constant-$k_{\rm f}$ mode was taken with a fixed final neutron energy $E_{\rm f}$ of 14.8 meV. In order to gain intensity, collimations were set as open - monochromator - $80'$ - sample - $80'$ - analyzer - $80'$ - detector. The energy resolution was about 2 meV because of loose collimations. A pyrolytic graphite filter was placed after the sample to suppress the higher order contaminations. We used a sample with a volume of approximately 2.5 cm$^3$. The sample was mounted in an ILL-type orange cryostat with its $a^*$- and $c^*$-axes in the scattering plane. The crystallographic parameters were determined as $a^*=1.6059$ 1/$\rm{\AA}$, $c^*=0.71513$ 1/$\rm{\AA}$ and $\cos\beta^*$=0.0967 at helium temperatures. 

In the previous neutron inelastic scattering measurements for TlCuCl$_3$ \cite{Oosawa2}, three excitations were observed in the energy range $E\leq15$ meV for the scans along $(h, 0, 0)$, $(h, 0, 1)$, $(1, 0, l)$, $(1.5, 0, l)$ and $(h,0,2h-1)$ with $1{\leq}h{\leq}1.5$ and $0{\leq}l{\leq}1$. From the temperature dependence of the excitation spectra, we concluded that the lowest dispersive excitation is of magnetic origin. Since the phonon excitation has been observed at almost the same energy in the isostructural KCuCl$_3$ \cite{Kato4}, we inferred that the highest dispersionless excitation can be attributed to the phonon excitation. However, the origin of the second excitation was unclear.

We first investigated the temperature variation of these excitations for ${\it{\pmb Q}}=(1.5,0,0)$. At $T=1.5$ K, three excitations were observed at $E\approx 3$, 7 and 12 meV. At room temperature, which is much higher than the temperatures corresponding to the excitation energies, the peak intensity of the lowest excitation decreases to the background level, while the intensities of the two higher excitations appear to be almost independent of temperature. This indicates that the two higher excitations are not intrinsic magnetic excitations. If the cause of the excitations is phonons, their intensities should increase with increasing temperature. Hence, it is too early to conclude that the two higher excitations are attributable only to phonons. 

In the present study, we focused on the lowest magnetic excitation, and investigated the precise dispersion relations along $(h, 0, 1)$, $(0, 0, l)$, $(h, 0, 2h+1)$ and $(h, 0, -2h+1.4)$ by constant-$\it{\pmb Q}$ energy scanning for $0\leq h\leq 0.5$ and $1\leq l\leq 2$ (see Fig. 2). The present scan area is closer to the origin of the reciprocal space than the previous one, so that the magnetic excitations can gain intensity due to the magnetic form factor. The dispersion relations along diagonal directions $(h, 0, 2h+1)$ and $(h, 0, -2h+1.4)$ are necessary to determine the exchange network uniquely within the data for the $a^*-c^*$ scattering plane.

Figure 3 shows the scan profiles for ${\it{\pmb Q}}=(h, 0, 1)$, $(0, 0, l)$,
$(h, 0, 2h+1)$ and $(h, 0, -2h+1.4)$ measured at $T=1.5$ K. A
well-defined single excitation can be observed in almost all
scans. The scan profiles were fitted with a Gaussian function to evaluate
the excitation energy, as shown by the solid lines in Fig. 3. The
horizontal bars in Fig. 3 denote the calculated resolution widths. Almost all peaks have widths equal to the resolution limit.

Figure 4 shows the constant-$\it{\pmb Q}$ scans for ${\it{\pmb Q}}=(0.1,0,1.2)$ measured at $T=1.5$ K and 80 K. A single excitation is clearly observed at $E=6.3$ meV and $T=1.5$ K. At $T=80$ K, the excitation spectrum broadens out to the background level. This indicates that the excitation can be attributed to a magnetic origin. 

The dispersion relation $\omega(\it{\pmb Q})$ obtained for the $a^*-c^*$
scattering plane is summarized in Fig. 5. It is evident that the
lowest excitation occurs at ${\it{\pmb Q}}=(0,0,1)$. However, since the
excitation energy is lower than 1 meV, we could not determine the
excitation energy due to the incoherent scattering and the low energy
resolution. In Fig. 5, we substituted the gap energy $\Delta = 0.65$
meV, which was evaluated from the previous magnetization
\cite{Shiramura1,Oosawa1} and ESR measurements \cite{Tanaka1}, for the excitation energy at ${\it{\pmb Q}}=(0,0,1)$. Recently Cavadini {\it et
al}. \cite{Cavadini4} investigated the magnetic excitations in
TlCuCl$_3$ by neutron inelastic scattering experiments. Their
experimental results along the $a^*$-and $c^*$-directions are in
agreement with our results.

Based on the present results, together with the previous ones, it is
evident that the periodicity of the magnetic excitation in TlCuCl$_3$
is the same as that of the nuclear reciprocal lattice along the
$a^*$-axis, but doubled along the $c^*$-axis, as observed in KCuCl$_3$
\cite{Kato1,Cavadini1}. Hence, it is deduced that the lowest
excitation occurs at ${\it{\pmb Q}}=(h, 0, l)$ with integer $h$ and odd $l$ 
in the $a^*-c^*$ plane. This is consistent with the results of recent neutron elastic scattering experiments in magnetic fields \cite{Tanaka2}. When a magnetic field $H$ is applied in the present system, the single excitation splits into three excitations, since the excitation should be a triplet
excitation. The lowest excitation energy decreases with increasing
magnetic field, and finally becomes zero at the critical field $H_{\rm
c}=\Delta/g\mu_{\rm B}$. For $H>H_{\rm c}$ the system can undergo
magnetic ordering due to three-dimensional (3D) interactions. Such
field-induced magnetic ordering has actually been observed by neutron
elastic scattering for $H\parallel b$ and $H>H_{\rm c}\approx 5.5$ T
\cite{Tanaka2}. Magnetic Bragg reflections were observed at ${\it{\pmb Q}}=(h, 0, l)$ with integer $h$ and odd $l$ in the $a^*-c^*$ plane, and are equivalent
to those for the lowest excitation at zero field.

There is a sharp contrast between the dispersion curves for ${\it{\pmb Q}}=(h, 0, 2h+1)$ and $(h, 0, -2h+1.4)$ which are roughly perpendicular to each other. The magnetic excitation is most dispersive along $(h, 0, 2h+1)$, which is parallel to the $(1, 0, -2)$ cleavage plane, and is less dispersive for $\it{\pmb Q}$ perpendicular to it. The dispersion curve for ${\it{\pmb Q}}=(h, 0, 2h+1)$ has a local minimum at $h=0.25$. This dispersion behavior is similar to that in KCuCl$_3$, which implies that the principal exchange pathways in TlCuCl$_3$ should be the same as those in KCuCl$_3$. However, the dispersion range in TlCuCl$_3$ ($0.65$ ${\rm meV}\leq\omega\leq 7.3$ ${\rm meV}$) is much larger than that in KCuCl$_3$ ($2.7$ ${\rm meV}\leq\omega\leq 5.0$ ${\rm meV}$). This suggests that the exchange interactions in TlCuCl$_3$ are much stronger than those in KCuCl$_3$. The details of the exchange interactions are evaluated in the next section. 

\section{ANALYSIS AND DISCUSSION}

The main point in the analysis of the experimental results is that the magnetic interactions in TlCuCl$_3$ are dominated
by the exchange between the two spins forming the planar $\rm
Cu_2Cl_6$ dimer (intradimer exchange), whereas exchange interactions
between other spin pairs (interdimer exchange) can be considered as weaker. The strong antiferromagnetic intradimer interaction is the
origin of the singlet ground state.

Since notable anisotropy effects have not been observed in the static measurements of TlCuCl$_3$ \cite{Oosawa1,Oosawa3}, the full magnetic interactions will
be described by the spin-$\frac{1}{2}$ Heisenberg model

\begin{eqnarray}
{\cal H}=\sum_{\langle i,j\rangle}J_{ij}\left({\it{\pmb S}}_{i}\,\cdot\,{\it{\pmb S}}_{j}\right)\ .
\end{eqnarray}

For the exchange interactions between spins, we will use the notation given in ref. \cite{Mueller}: the main intradimer exchange is denoted
as $J$. The exchange interaction per bond between spins in dimers
separated by a lattice vector $l{\it{\pmb a}} + m{\it{\pmb b}} + n{\it{\pmb c}}$
is denoted as the exchange energy $J_{lmn}$ for pairs of spins at
equivalent positions in their respective dimer and as $J_{lmn}'$ for
spins at inequivalent positions. Finite values of interdimer
interactions $J_{\dots}$ will be considered for the following exchange
paths only: $ (lmn) = (100), (200), (1 \frac{1}{2} \frac{1}{2}), (0
\frac{1}{2} \frac{1}{2})$ and $(201)$. In the following we present an
analysis of the lowest elementary triplet excitation based on this
picture, in three steps.

(i) The simplest picture is to assume unperturbed propagation of the
excitation of one dimer from the singlet to triplet state, neglecting the
excitation of further triplets during propagation. This leads to the
basic dispersion

\begin{eqnarray}
\omega_{\pm}({\it{\pmb Q}}) &=& J + \delta \omega_{\pm}^{(1)}({\it{\pmb Q}})\ ,\\
\delta \omega_{\pm}^{(1)}({\it{\pmb Q}}) &=&  
    [J_{(100)}^{\rm eff}{\cos}(2{\pi}h) +
     J_{(200)}^{\rm eff}{\cos}(4{\pi}h) +
     J_{(201)}^{\rm eff} {\cos}\{ 2{\pi}(2h+l)\}]  \nonumber \\ 
& & {\pm} 2 
[J_{(1\frac{1}{2}\frac{1}{2})}^{\rm eff} {\cos}\{{\pi}(2h+l)\}{\cos}({\pi}k) + 
 J_{(0\frac{1}{2}\frac{1}{2})}^{\rm eff} {\cos}({\pi}k){\cos}({\pi}l)\ .
\end{eqnarray}
 
At this level, only certain combinations of the exchange interactions
enter the dispersion law: these are known as effective
dimer interactions:

\begin{eqnarray}
J^{\rm eff}_{(100)} &= & \frac{1}{2}\left(2J_{(100)}-J'_{(100)}\right)\ , \nonumber \\
J^{\rm eff}_{(200)} &= & \frac{1}{2}\left(2J_{(200)}-J'_{(200)}\right)\ , \nonumber \\ 
J^{\rm eff}_{(1\frac{1}{2}\frac{1}{2})} &= & \frac{1}{2}\left(J_{(1\frac{1}{2}\frac{1}{2})}-J'_{(1\frac{1}{2}\frac{1}{2})}\right)\ , \nonumber \\
J^{\rm eff}_{(0\frac{1}{2}\frac{1}{2})} &= & \frac{1}{2}\left(J_{(0\frac{1}{2}\frac{1}{2})}-J'_{(0\frac{1}{2}\frac{1}{2})}\right)\ ,\nonumber \\
J^{\rm eff}_{(201)} &= & -\frac{1}{2}J'_{(201)}\ .
\end{eqnarray}

The factor in front of the individual exchange constants $J$ and $J'$
results from the number of identical exchange paths for the same
lattice vector between dimers; it is one of 1 (identical exchange on
two legs or on two diagonals), $\frac{1}{2}$ (exchange on one
diagonal only whereas there is no exchange path for the other diagonal)
or 0 (the exchange path does not contribute).
Dispersion relation (2) is valid in the limit $J_{lmn}, J'_{lmn}
\ll J$. 

Since there are two different dimers per chemical unit cell, the
dispersion law has two branches, distinguished by $\pm$ in eq.(2).
However, the structure factor at the zeroth order
is of the form
\begin{eqnarray}
S({\it{\pmb Q}},\omega_{\pm})\sim\left(\sin{\frac{{\it{\pmb Q}}\cdot{\it{\pmb R}}_1}{2}} \pm
             \sin{\frac{{\it{\pmb Q}}\cdot{\it{\pmb R}}_2}{2}}\right)^2\ ,
\end{eqnarray}
where ${\it{\pmb R}}_1=0.47{\it{\pmb{a}}}+0.10{\it{\pmb{b}}}+0.31{\it{\pmb{c}}}$ and
${\it{\pmb R}}_2=0.47{\it{\pmb{a}}}-0.10{\it{\pmb{b}}}+0.31{\it{\pmb{c}}}$ denote the spin
separations in the Cu$_2$Cl$_6$ dimers located at the corner and at the center
of the unit cell in the $b-c$ plane. Under the present experimental
condition, {\it i.e.}, $\it{\pmb Q}$ in the $a^*-c^*$ plane, the
$\omega_{+}({\it{\pmb Q}})$ branch gives the only nonvanishing
contribution. Hence, we can assume that the observed single excitation
corresponds to the $\omega_{+}({\it{\pmb Q}})$ branch.

(ii) The simple approach (i) gives a qualitatively correct picture,
which confirms the assumption of one dominating intradimer exchange
interaction, quantitatively. However, it is not sufficient. The
standard way to improve on this has been to treat the intermediate
excitation of two or more triplets in an RPA-like approximation
\cite{Cavadini1,Leuenberger}. This approximation continues to treat
dimers as units and results in the dispersion law 
\begin{eqnarray}
\omega_{\pm}({\it{\pmb Q}})=\sqrt{J^2+ 2J \delta \omega_{\pm}^{(1)}({\it{\pmb Q}})}.
\end{eqnarray} 
This result depends only on the effective interactions $J^{\rm
eff}_{lmn}$; the dispersion law does not distinguish between the
contributions to the interdimer interactions from the different
exchange paths.

The measured dispersions were fitted to eq. (6) using all the
effective exchange constants given in eq. (4). The results are
presented in Table II. 
 
(iii) A systematic approach to improve on the one triplet dispersion
law of eq. (2), is to expand the energy of the elementary triplet
excitation order by order in the interdimer exchange interactions.
The method has been described in \cite{Mueller}, and it has been found that, from the second order, new terms which are not present in the expansion of the square root of eq. (6) appear. To determine individual exchange couplings, we have implemented the cluster expansion technique \cite{Mueller,gelfand1,gelfand2} to allow us to calculate one magnon dispersion relations up to the sixth order. For tractability, only exchange paths along (100), (201) and $(1 \frac{1}{2} \frac{1}{2})$ were taken into account, whereas the small (as evident from approach (ii)) interactions along (200) and $(0 \frac{1}{2} \frac{1}{2})$ were disregarded.
Due to the complex interaction structure, it was not possible to take
into account only topologically different clusters, and we had to consider
all embeddings, {\it i.e.}, all the different ways of setting a given cluster on
the lattice, up to the desired order. This leads to an exponentially
growing number of cluster from order to order. In the present case, we perform calculations for the sixth order treating 18084 colored clusters and 673826
embeddings.

The series expansion gives the dispersion relation in the form
\begin{equation}\label{eq:seromega}
\omega({\it{\pmb Q}})/\!J = 1 + \sum_{p_1\ge 0,\ldots,p_5 \ge 0}\sum_{(n_h,n_k,n_l)} C_{n_h,n_k,n_l}^{p_1,\ldots,p_5}\prod_{j=1}^5 \alpha_j^{p_j}\cos\left[\pi(n_h h+n_l l)\right]\cos(\pi n_k k),
\end{equation}
where $\alpha_i$'s are the individual exchange interactions in
units of $J$:
\begin{eqnarray}
\alpha_1 = J^{\prime}_{(201)}/\!J, \ \ 
\alpha_2 = J_{(100)}/\!J, \ \ 
\alpha_3 = J^{\prime}_{(100)}/\!J, \ \ 
\alpha_4 = J_{(1\frac 1 2 \frac 1 2)}/\!J, \ \ 
\alpha_5 = J^{\prime}_{(1 \frac 1 2 \frac 1 2)}/\!J.
\end{eqnarray}
Leading terms in (\ref{eq:seromega}) can be combined to an expansion
of a square root, as discussed in \cite{Mueller}. We have calculated
the coefficients $C_{n_h,n_k,n_l}^{p_1,\ldots,p_5}$ explicitly up to the fourth
order, {\it i.e.}, $\sum_{j=1}^5 p_j \le 4$; these coefficients are
available up on request. Up to the sixth order, we are able to obtain the
dispersion numerically for a given parameter set $\{\alpha_i\}$. In order to demonstrate the convergence of the cluster expansion we show in Fig. 6 the results for increasing orders in the dimer expansion for the direction ${\it{\pmb Q}}=(h,0,1)$. 

For the analysis of the data, we proceeded in two steps. We first
performed a least-squares fit to the measured data using the
analytically known dispersion to the fourth order with the intradimer interaction $J=5.68$ meV, which was obtained in (ii). Guided by the result of this fit, we performed calculations to the sixth order. The parameters for
the best sixth order calculation are given in Table III. The solid lines in Fig. 5 indicate the calculated results. The experimental dispersion curves can be reproduced well by the present calculation.

The results for the effective exchange constants are well defined and imply, in particular, a significant increase in the value of $J^{\rm eff}_{(1 \frac{1}{2} \frac{1}{2})}$, when compared to the RPA-like approach (ii) above. The individual exchange constants are well defined for $(lmn)=(100)$ and $(201)$. For $(lmn) = (1 \frac{1}{2} \frac{1}{2})$ the individual exchange constants are determined less reliably. A fit of comparable accuracy is obtained for $J_{(1 \frac{1}{2} \frac{1}{2})} = 0.23 \;{\rm meV}$, $ J^{\prime}_{(1 \frac{1}{2} \frac{1}{2})}=-1.37 \; {\rm meV}$ and minor changes in the remaining parameters. Similar to the situation in $\rm KCuCl_3$, the neighboring dimers couple magnetically along the chain and in the $(1, 0, -2)$ plane. The most important interdimer interaction is the diagonal $J^{\prime}_{(201)}$ interaction, which is about half of the intradimer interaction.

\section{CONCLUSIONS}
We have presented the results of neutron inelastic scattering for the
spin gap system TlCuCl$_3$. Well-defined magnetic excitation spectra
were observed in the $a^*-c^*$ plane. The dispersion relations of the
magnetic excitations in TlCuCl$_3$ were determined as shown in Fig. 5,
and were analyzed by the cluster series expansion. The individual
exchange interactions, which could not be obtained from the effective
dimer approximation, were evaluated, as shown in Table III. TlCuCl$_3$ is a strongly coupled spin-dimer system, in contrast to KCuCl$_3$
which is characterized as a weakly coupled spin-dimer system. The
analysis of the individual exchange constants shows that the ladder
system in $\rm TlCuCl_3$, similar to $\rm KCuCl_3$, is much closer to
an alternating spin chain than has been previously believed.

\acknowledgments
The authors would like to thank N. Cavadini for sending them reference \cite{Cavadini4} prior to its publication. This work was supported by Toray Science Foundation and a Grant-in-Aid for Scientific Research on Priority Areas (B) from the Ministry of Education, Culture, Sports, Science and Technology of Japan. A. O. was supported by the Research Fellowships of the Japan Society for the Promotion of Science for Young Scientists. \par

\begin{figure}[ht]
\vspace*{3cm}
 \epsfxsize=150mm
  \centerline{\epsfbox{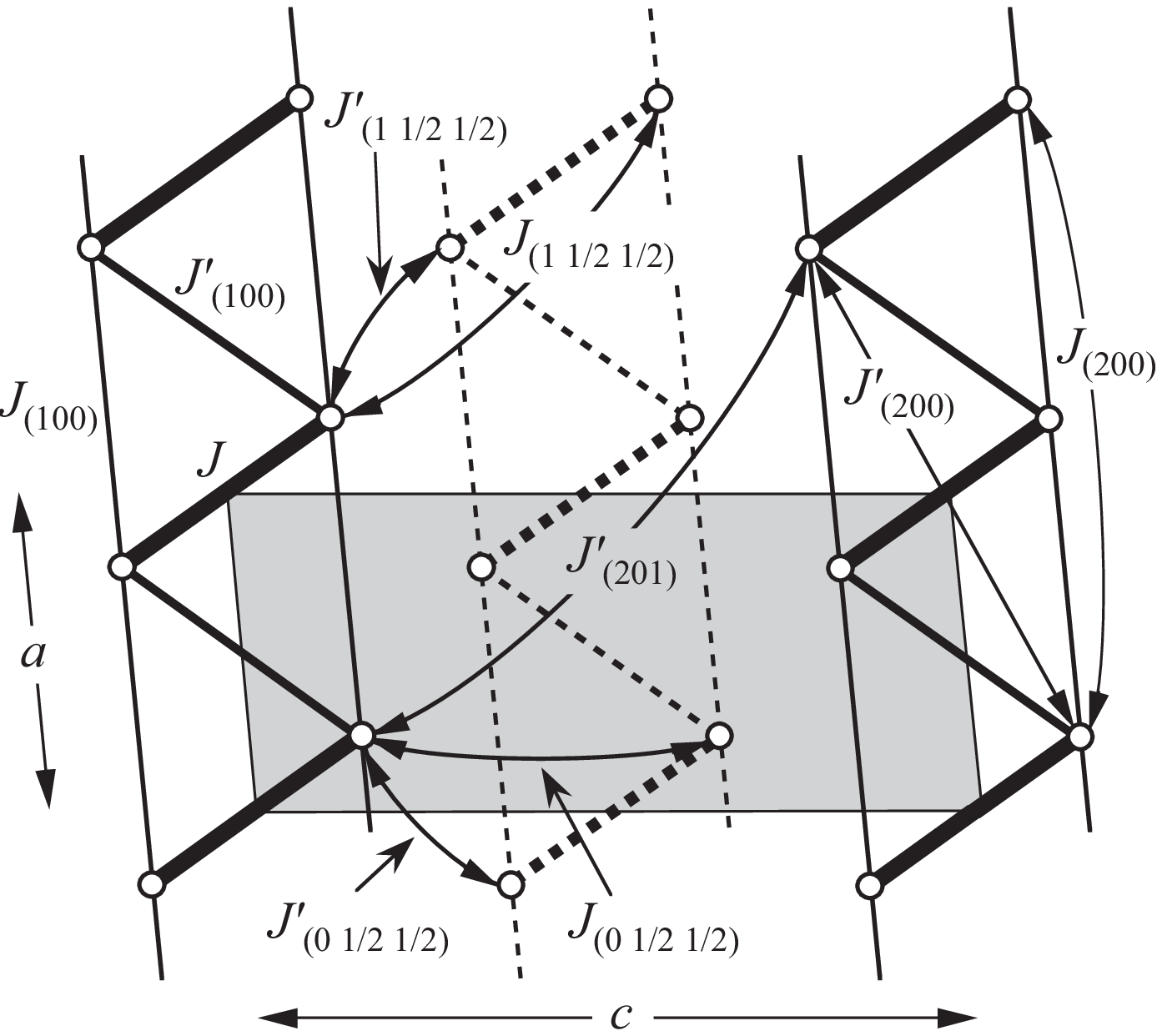}}
\vspace{1cm}
\caption{Projection of Cu$^{2+}$ ions with spin-$\frac{1}{2}$ on the $a-c$ plane and the exchange interactions. The double chains located at the corner and the center of the chemical unit cell in the $b-c$ plane are represented by solid and dashed lines, respectively. The shaded area is the chemical unit cell in the $a-c$ plane. }
\label{Fig.1}
\end{figure}

\newpage
 
\begin{figure}[ht]
\vspace*{3cm}
 \epsfxsize=100mm
  \centerline{\epsfbox{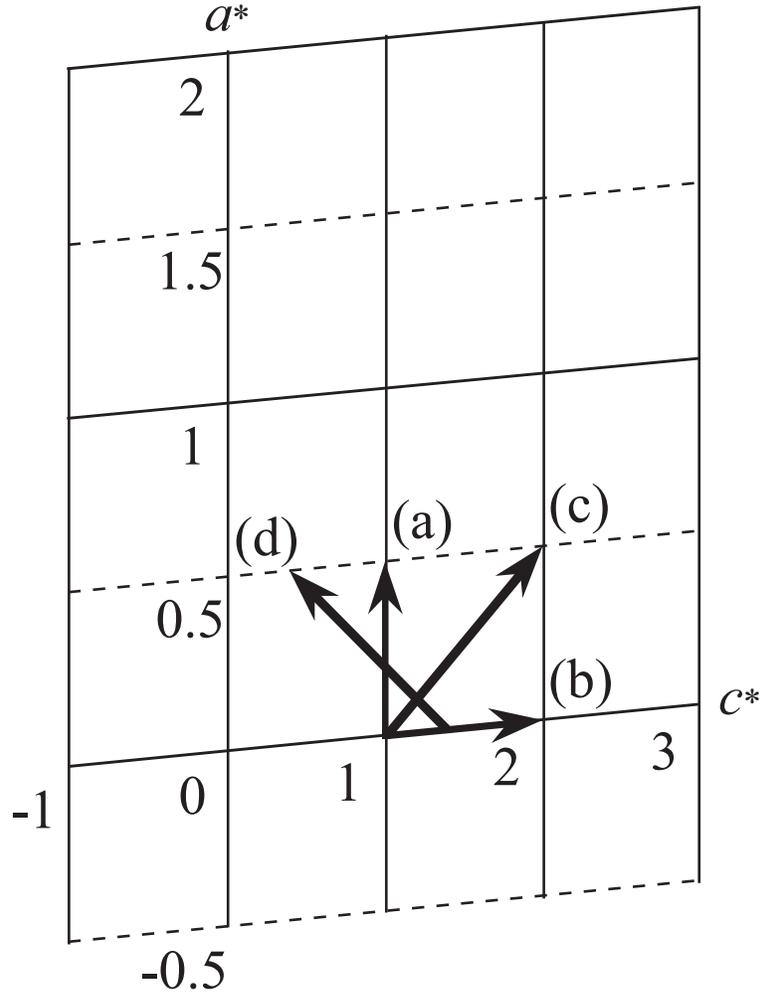}}
\vspace{1cm}
\caption{Scanning directions for $\it{\pmb Q}$ along (a) $(h, 0, 1)$, (b) $(0, 0, l)$, (c) $(h, 0, 2h+1)$ and (d) $(h, 0, -2h+1.4)$.}
\label{Fig.2}
\end{figure}

\newpage

\begin{figure}[ht]
\begin{minipage}{7.5cm}
 \epsfxsize=55mm
  \centerline{\epsfbox{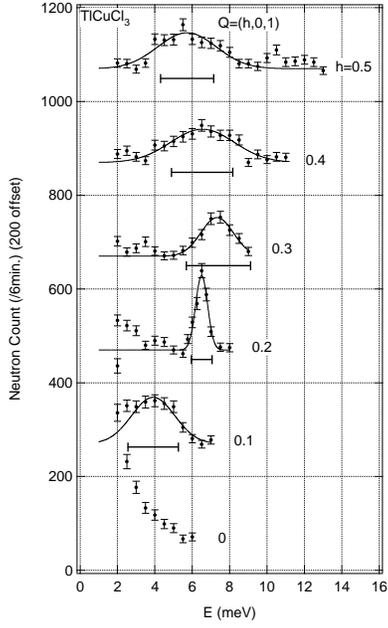}}
\begin{center}
(a)
\end{center}
\end{minipage}
\begin{minipage}{7.5cm}
 \epsfxsize=55mm
  \centerline{\epsfbox{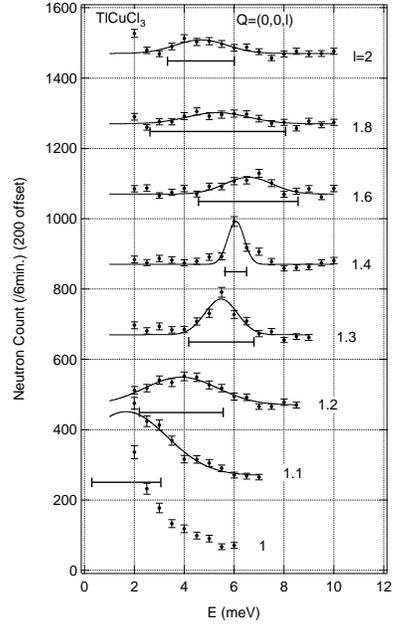}}
\begin{center}
(b)
\end{center}
\end{minipage} \par
\begin{minipage}{7.5cm}
 \epsfxsize=55mm
  \centerline{\epsfbox{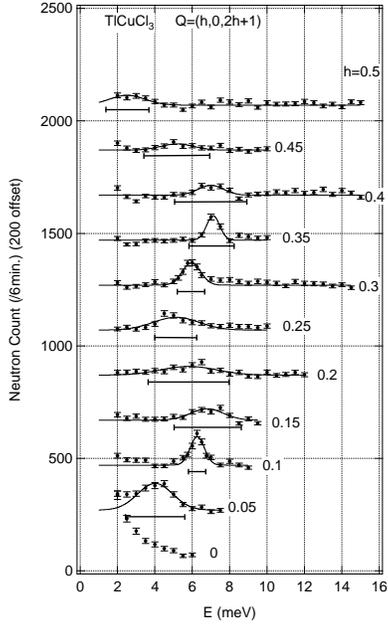}}
\begin{center}
(c)
\end{center}
\end{minipage}
\begin{minipage}{7.5cm}
 \epsfxsize=55mm
  \centerline{\epsfbox{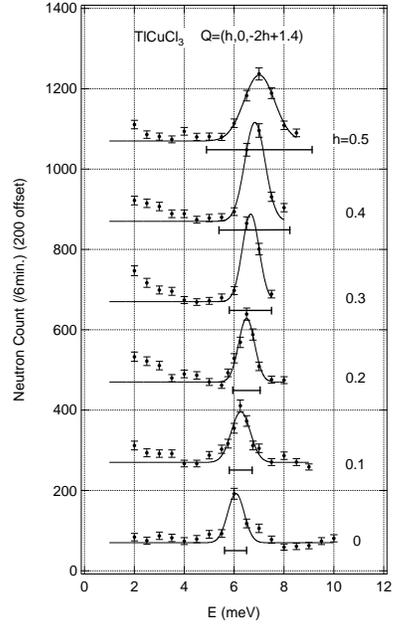}}
\begin{center}
(d)
\end{center}
\end{minipage}
\vspace{1cm}
\caption{Profiles of the constant-${\vec Q}$ energy scans in TlCuCl$_3$ for $\it{\pmb Q}$ along (a) $(h, 0, 1)$, (b) $(0, 0, l)$, (c) $(h, 0, 2h+1)$ and (d) $(h, 0, -2h+1.4)$ with $0\leq h\leq 0.5$ and $1\leq l\leq 2$. The solid lines are fit using a Gaussian function. The horizontal error bars indicate the calculated resolution widths.} 
\label{Fig.3}
\end{figure}

\newpage

\begin{figure}[ht]
\vspace*{3cm}
\begin{center}
 \epsfxsize=150mm
  \centerline{\epsfbox{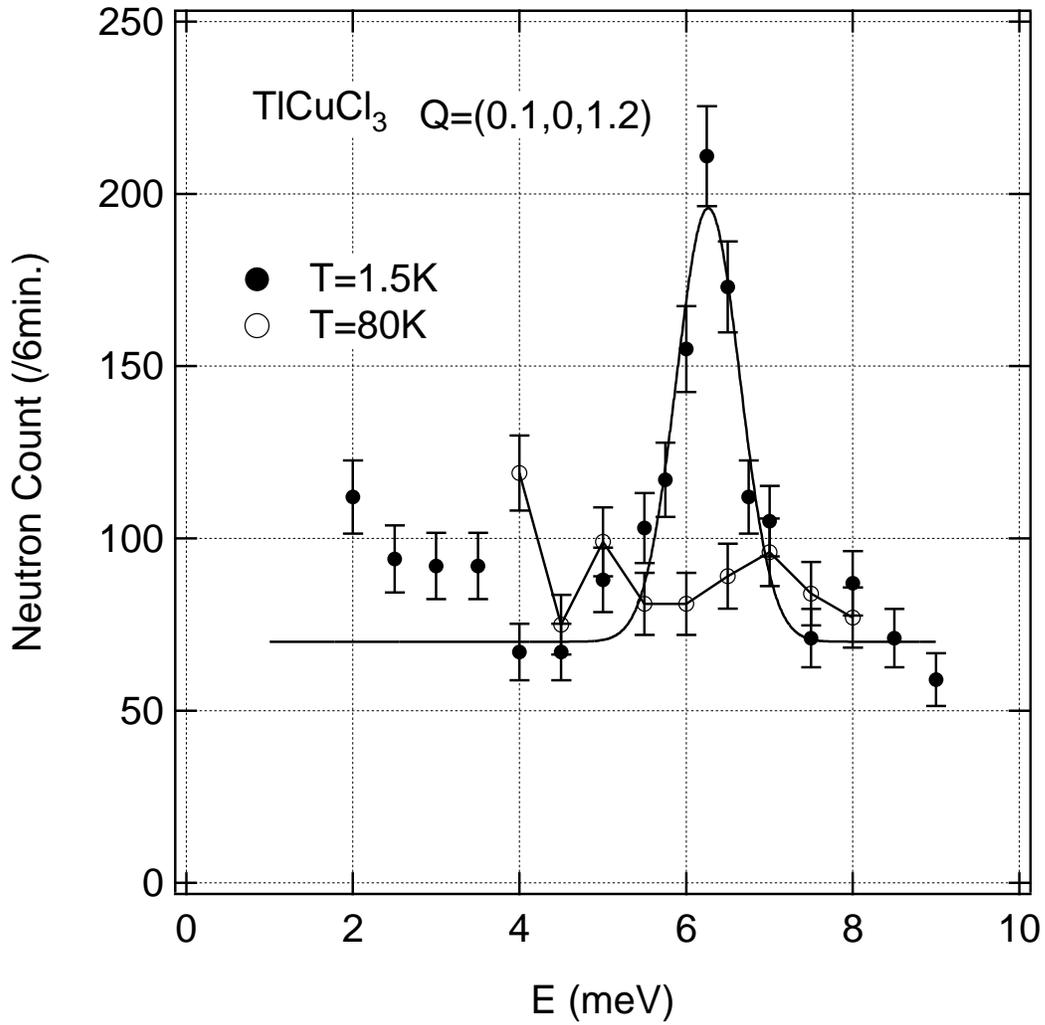}}
\end{center}
\vspace{1cm}
\caption{Constant-$\it{\pmb Q}$ energy scans in TlCuCl$_3$ at ${\it{\pmb Q}}=(0.1,0,1.2)$ for $T=1.5$ K and 80 K. The solid line for $T=1.5$ K is a Gaussian fit.}
\label{Fig.4}
\end{figure}

\newpage

\begin{figure}[ht]
\vspace*{3cm}
\begin{minipage}{7.5cm}
\begin{center}
 \epsfxsize=70mm
  \centerline{\epsfbox{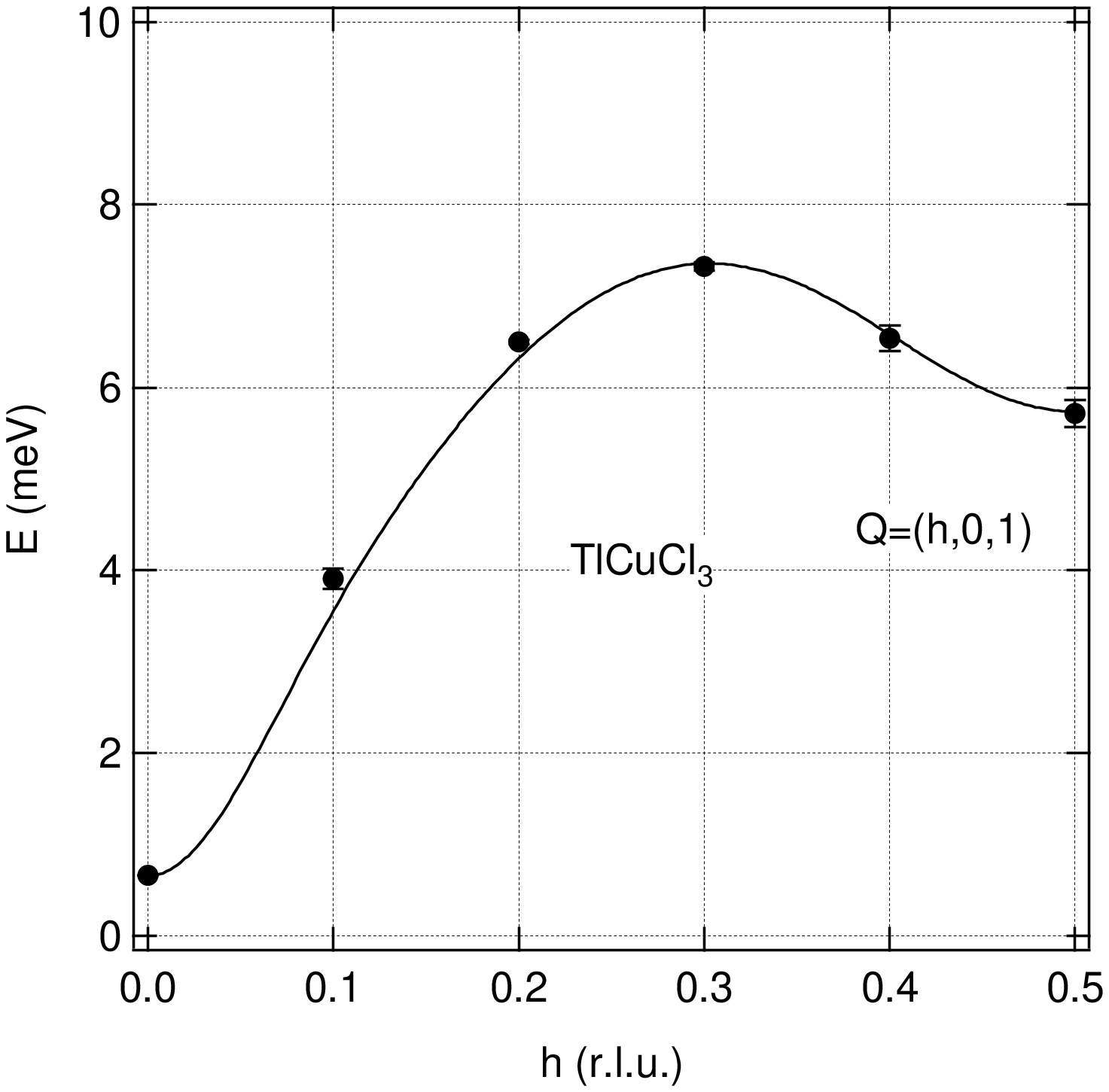}}
(a)
\end{center}
\end{minipage}
\begin{minipage}{7.5cm}
\begin{center}
 \epsfxsize=70mm
  \centerline{\epsfbox{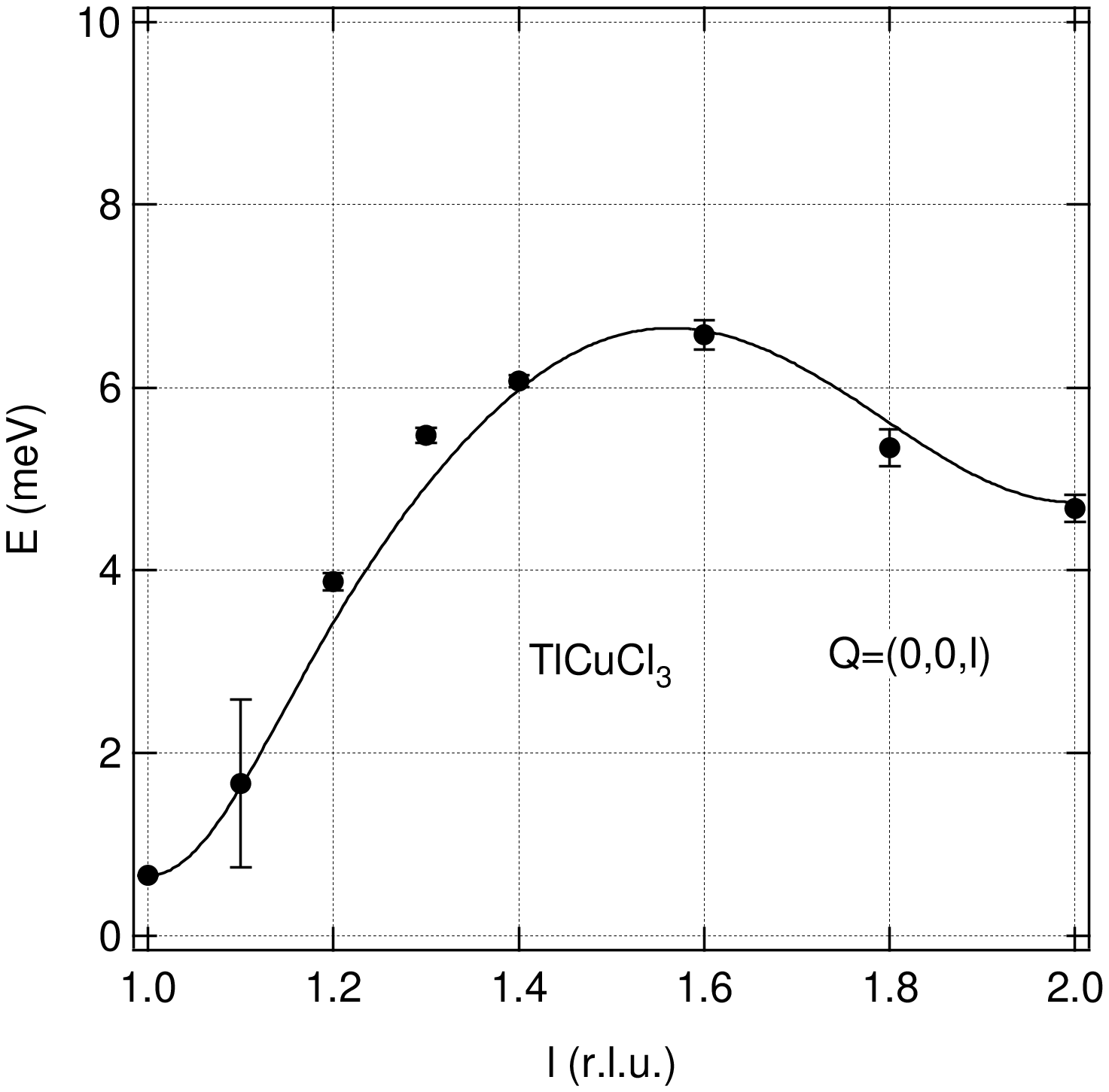}}
(b)
\end{center}
\end{minipage}\par
\begin{minipage}{7.5cm}
\begin{center}
 \epsfxsize=70mm
  \centerline{\epsfbox{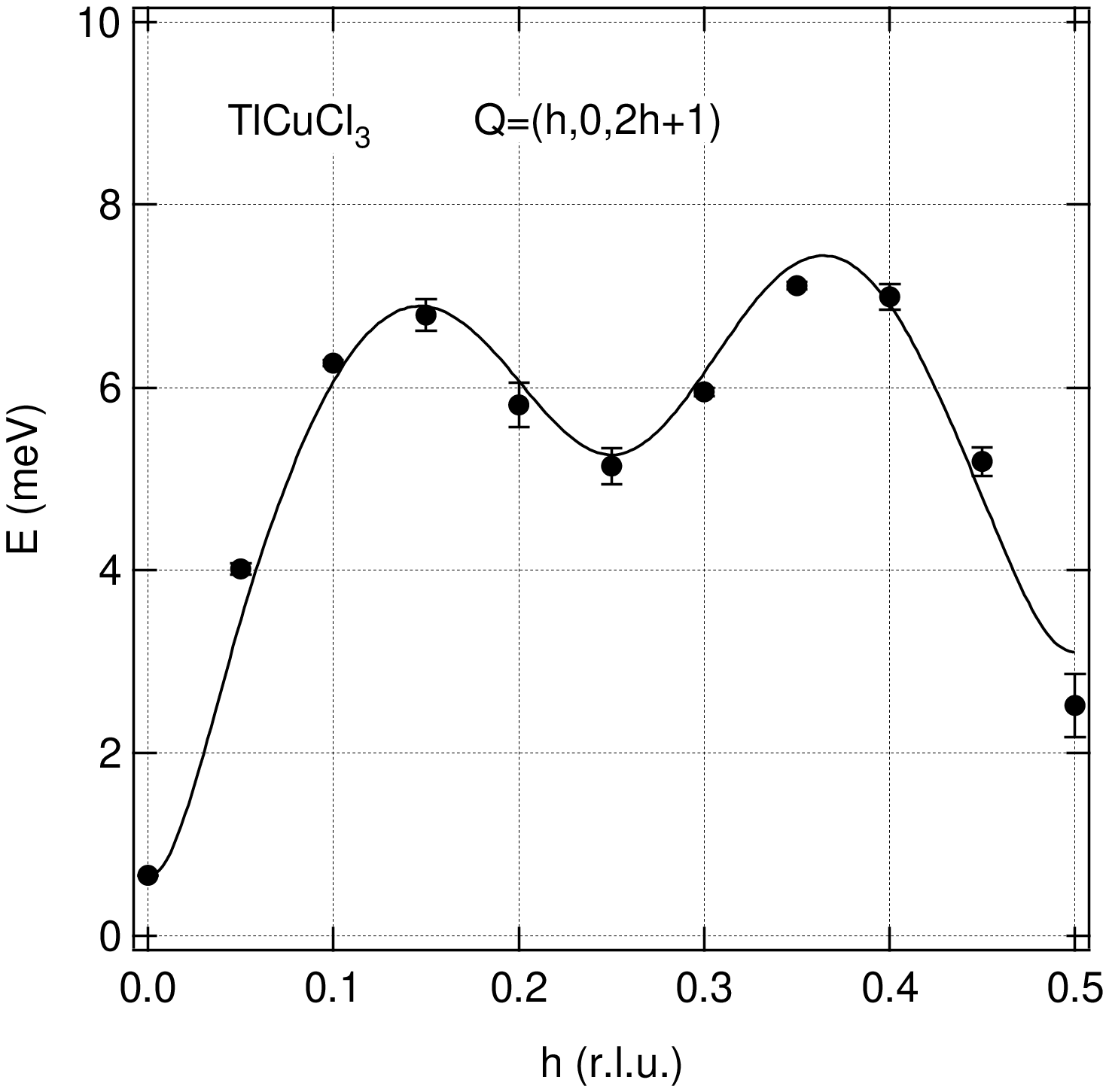}}
(c)
\end{center}
\end{minipage}
\begin{minipage}{7.5cm}
\begin{center}
 \epsfxsize=70mm
  \centerline{\epsfbox{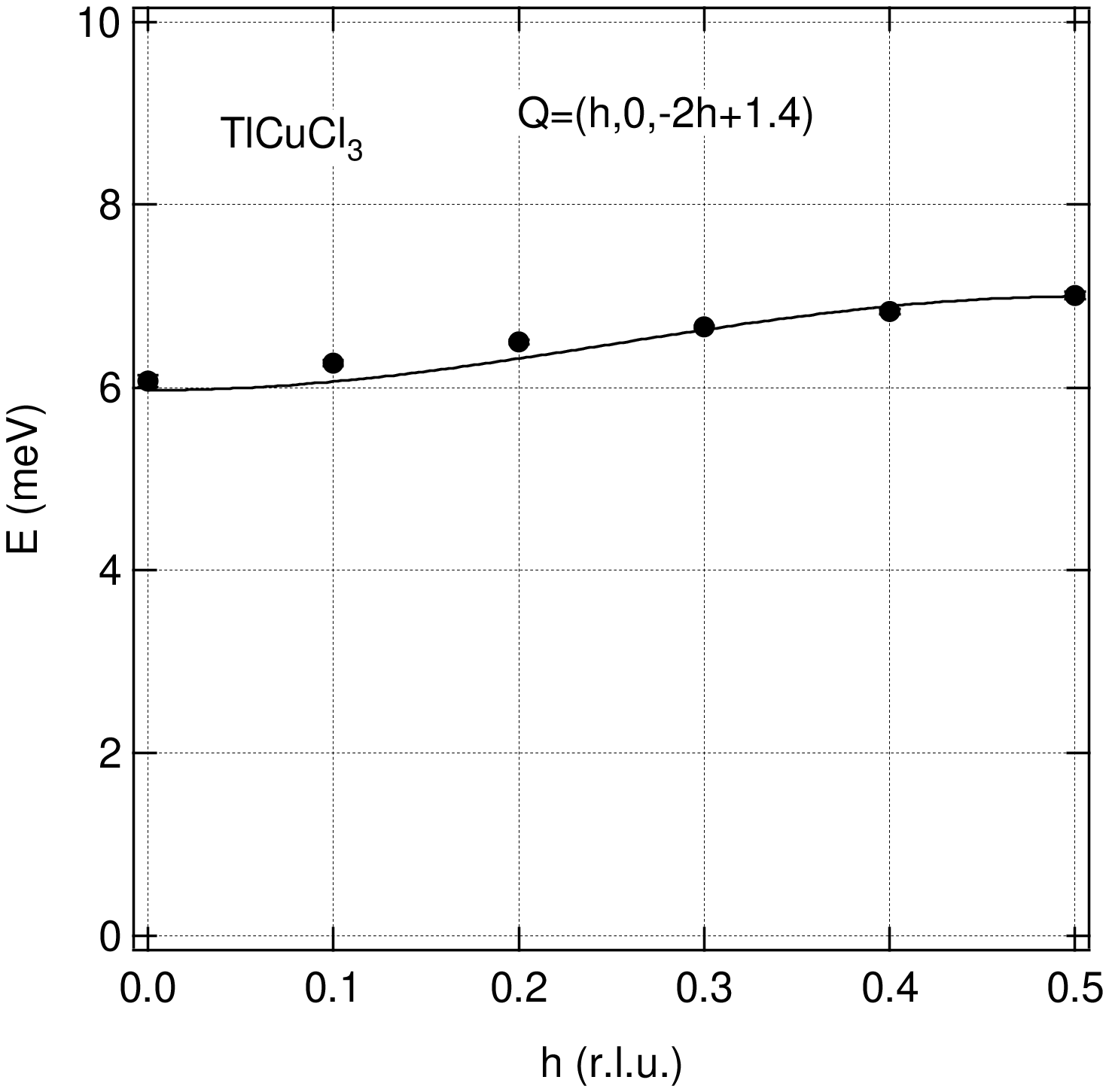}}
(d)
\end{center}
\end{minipage}
\vspace{1cm}
\caption{Dispersion relations $\omega(\it{\pmb Q})$ in TlCuCl$_3$ for $\it{\pmb Q}$ along (a) $(h, 0, 1)$, (b) $(0, 0, l)$, (c) $(h, 0, 2h+1)$ and (d) $(h, 0, -2h+1.4)$. Solid lines are the dispersion curves calculated by cluster series expansion to the sixth order using the exchange constants in Table III.} 
\label{Fig.5}
\end{figure}

\newpage

\begin{figure}[ht]
\vspace*{3cm}
\begin{center}
 \epsfxsize=150mm
  \centerline{\epsfbox{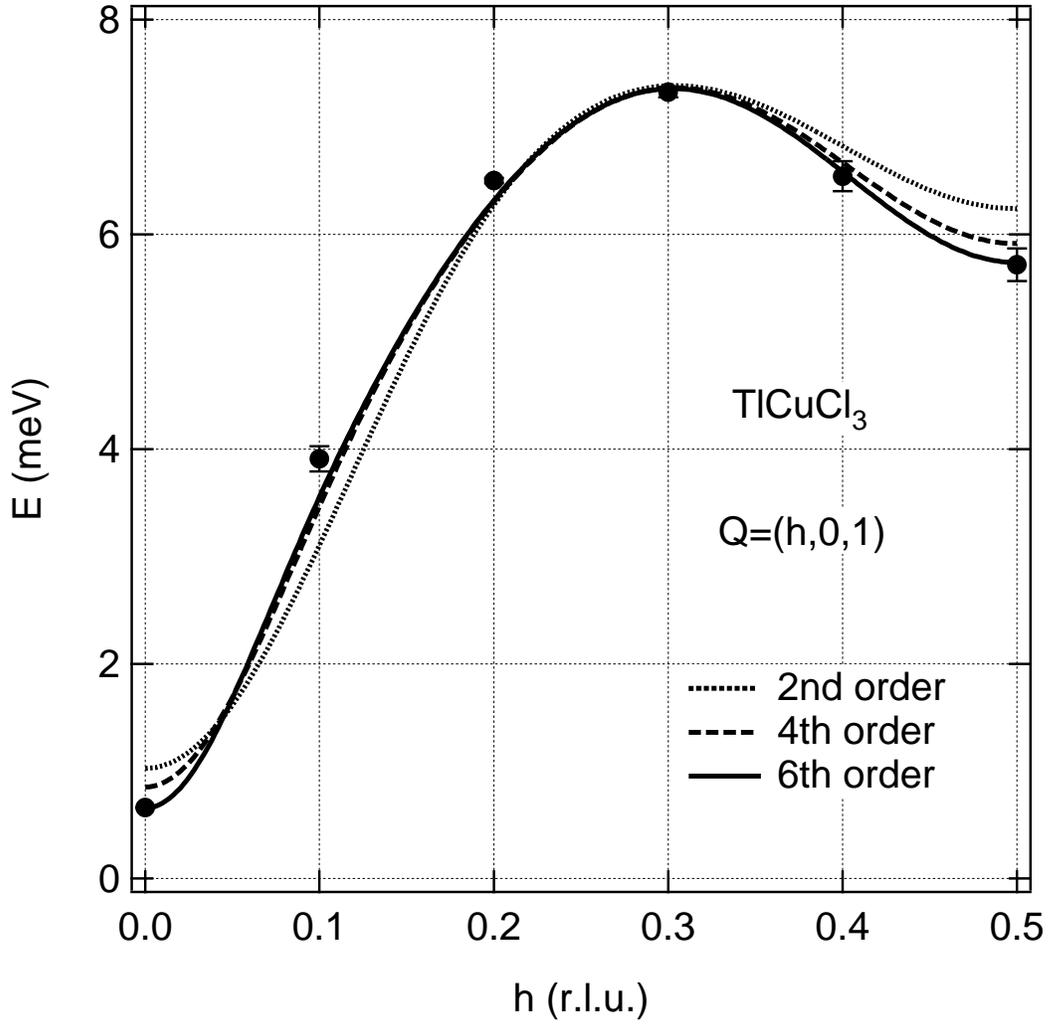}}
\end{center}
\vspace{1cm}
\caption{The convergence of the cluster expansion with increasing orders in the dimer expansion for ${\it{\pmb Q}} = (h,0,1)$. Dotted, dashed and solid lines denote the results for the second, fourth and sixth orders with the exchange constants in Table III. Closed circles are experimental data.}
\label{Fig.6}
\end{figure}

\newpage

\begin{table}
\caption{The lattice constants $a, b, c$ and $\beta$ at room temperature, the critical fields $H_{\rm c}=\Delta/g\mu_{\rm B}$ and the saturation fields $H_{\rm s}$ for KCuCl$_3$ and TlCuCl$_3$.}
\label{table1}
\begin{tabular}{cccccc} 
  & KCuCl$_3$ & ref. & & TlCuCl$_3$ & ref.\\ 
\tableline
$a$ $(\rm{\AA})$        & 4.029 & \cite{Willett} & & 3.982 & \cite{Tanaka2}\\
$b$ $(\rm{\AA})$        & 13.785 & \cite{Willett} & & 14.144 & \cite{Tanaka2}\\
$c$ $(\rm{\AA})$ & 8.736        & \cite{Willett} & & 8.890 & \cite{Tanaka2}\\ 
$\beta$ & 97.33$^{\circ}$ & \cite{Willett} & & 96.32$^{\circ}$ & \cite{Tanaka2}\\ 
$(g/2)H_{\rm c}$ (T) & 23.1 & \cite{Shiramura1} & & 5.7 & \cite{Oosawa3}\\
$(g/2)H_{\rm s}$ (T) & 54.5 & \cite{Tatani} & & $\sim 100$ & \cite{Tatani}\\  
\end{tabular}
\end{table}

\begin{table}
\caption{The intradimer interaction $J$ and the effective interdimer interactions $J^{\rm eff}_{(lmn)}$ in KCuCl$_3$ and TlCuCl$_3$. All energies are in units of meV.}
\label{table2}
\begin{tabular}{ccc}
$J^{\rm eff}$ [meV] & KCuCl$_3$ \cite{Cavadini1,Kato3} & TlCuCl$_3$ \\
\tableline
$J$ & 4.34 & 5.68 \\
$J^{\rm eff}_{(100)}$ & $-0.21$ & $-0.46$ \\
$J^{\rm eff}_{(200)}$ & 0.03 & 0.05 \\
$J^{\rm eff}_{(1\frac{1}{2}\frac{1}{2})}$ & 0.28 & 0.49 \\
$J^{\rm eff}_{(0\frac{1}{2}\frac{1}{2})}$ & $-0.003$ & $-0.06$ \\
$J^{\rm eff}_{(201)}$ & $-0.45$ & $-1.53$ \\
\end{tabular}
\end{table}

\begin{table}
\caption{Interdimer exchange interactions in TlCuCl$_3$ determined by the cluster series expansion with the dimer interaction $J=5.68$ meV. All energies are in units of meV. AF and F denote antiferromagnetic and ferromagnetic exchange interactions, respectively.}
\label{table3}
\begin{tabular}{ccc}
$J_{(100)}= 0.34 $ (AF), \  $J'_{(100)}= 1.70$ (AF) & $\Rightarrow$ & 
$J^{\rm eff}_{(100)}= -0.51$ \\
$J_{(1\frac{1}{2}\frac{1}{2})} = 0.91 $ (AF), \  
$J'_{(1\frac{1}{2}\frac{1}{2})} = -0.57 $ (F) & $\Rightarrow$ & 
$J^{\rm eff}_{(1\frac{1}{2}\frac{1}{2})} = 0.74 $ \\
$J'_{(201)} = 2.56 $ (AF) & $\Rightarrow$ &
$J^{\rm eff}_{(201)} = -1.28 $ \\
\end{tabular}
\end{table}

\end{document}